# DIAGNOSTICS NEEDS FOR BEAM–BEAM STUDIES AND OPTIMIZATION

R. Giachino, CERN, Geneva, Switzerland

*Abstract*

During the recent years of LHC operation, we analysed the situation of beam instrumentation and the need to optimize it for beam–beam studies. The most important beam instrumentation devices will be highlighted and modifications or optimizations will be suggested. A complete wish list will be presented to make sure we will be ready after LS1 (Long Shutdown 1) to study the beam–beam effect in a more complete way.

## INTRODUCTION

The LHC beam–beam studies during the last three years have been very fruitful, mainly thanks to machine availability and the preparation of the beams at the injectors. The help of experts in the SPS and the CPS complex to obtain very intense bunch with low emittance were the important prerequisites to study the head-on beam–beam phenomena in the LHC. A full set of measurements has taken place in the LHC, for example to study beam–beam losses for decreasing crossing angle [1], to overcome instabilities due to the loss of Landau damping [2, 3] and to have a detailed measurement of the chromaticity and its possible bunch-by-bunch differences.

During these studies, we have heavily used the beam instrumentation devices and we could see that some important instrumentation was missing or not yet operational for beam–beam granularity measurements.

## MOTIVATION

Beam instrumentation is extremely important for beam–beam studies, e.g. in dedicated study of instabilities caused by beam–beam effects and data analysis to understand the phenomena. An ambitious target would be to have a bunch-by-bunch measurement of tune, chromaticity, intensity and transverse emittance. In order to control the beam stability we need to measure the turn-by-turn bunch position to determine the rise time and the frequency of the instability. This would allow us to measure and determine which mode is unstable, which instability type is present, as well as whether it is single/coupled bunch, single/coupled beam. An increased control of the stability would also help to understand the cause of the instability, e.g. impedance, electron cloud, beam–beam effect or a combination of the above phenomena. Another important achievement toward a better stability control would be to determine which mode is present. The ADT future development might allow us to measure the two beams synchronized in time (at the turn level). The new head tail monitor could give access to intra-bunch motion and a reliable, continuous chromaticity measurement. Finally, an interesting study could be to look into combined beam–beam/impedance modes (Fig. 1).

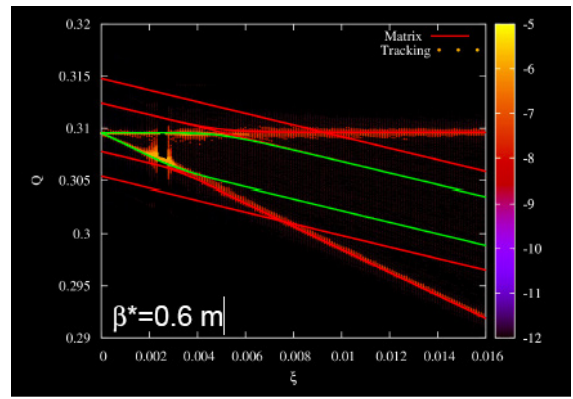

Figure 1: Instability resulting from coupling beam–beam/impedance modes [4].

Another important reason for bunch-by-bunch measurement is to be able to characterize bunch parameters from the beam–beam perspective and correlate them to luminosity lifetime degradation. If we compare this to simulation and try to understand the process, we could reduce the trial and error method and improve optimization of the luminosity lifetime. A working point optimization gave a 15–20% reduced emittance growth over the first 15 minutes in HEP stores in Tevatron Run II as the beam moved away from 5th and 12th order resonance [5]. The parameter space to scan is large, and tune, chromaticity, octupoles and transverse damper settings are the most important candidates to improve beam stability and luminosity lifetime (Fig. 2).

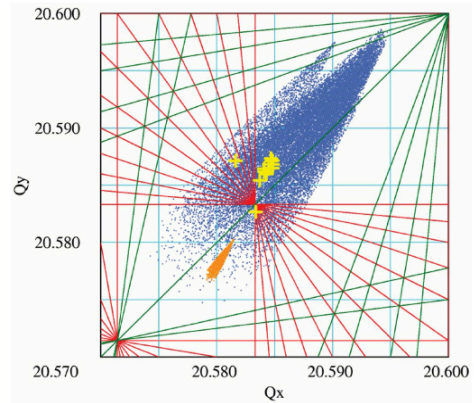

Figure 2: Tevatron tune footprint.

During regular operation, the rate of change of beam parameters is slow and therefore measurements with a sampling of a few seconds to one minute would be suitable.

In the squeeze or in the collision beam processes, very fast instabilities can develop. We need a triggered measurement lasting for a few seconds to one minute to record these instabilities. A snapshot of the machine parameters and bunch-by-bunch measurement covering the full collision beam process (of length ~60 s) would be the adequate tool to observe and understand the nature of the instabilities.

If instabilities due to beam–beam effects provoke a beam dump, the only observable signal will be available from the post mortem system (data from different measurement systems is recorded on circular buffers, frozen and exported into a beam dump event in the database). A possible improvement will be to extend the buffer to 1 s.

We stress the need for:
- High resolution, as the instabilities are fast (rise times can be well below 200 turns).
- Large data buffers, to acquire measurements for all bunches for a maximum number of turns.
- A triggering system, to be able to freeze buffers and acquire data from 1 to 2 seconds before the instabilities are detected (for example for tune measurements, BBQ, beam losses, BLMs, etc.), and maximize the chance that useful data to understand the instabilities are recorded when they occur.

## OBSERVATION OF INSTABILITIES

In 2012, the optimization of the machine performance was done by using small emittance beams, small β*, tight collimator settings (leading to large impedance) and high intensity per bunch. This brought along instabilities (Fig. 3), which started to limit the LHC efficiency and physics production. If we can measure tune and chromaticity for specific bunches we will be able to understand better the origin of these instabilities and possibly cure them [6].

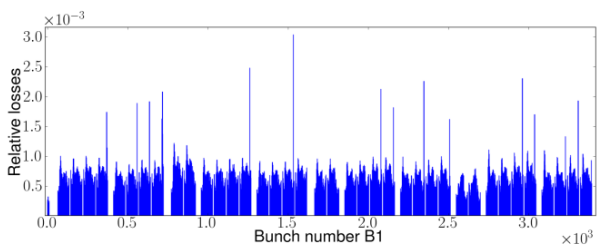

Figure 3: Bunch-by-bunch relative losses during the end of squeeze instability: only some bunches are affected by the instability and suffer losses that are more than twice those suffered by other bunches.

Instabilities in 2012 were essentially observed by BLM, BBQ, MIM (multiband-instability-monitor) and transverse damper beam position signals. The new head–tail monitors will enable measurement of higher frequencies, and thus to look inside the bunch and understand the type of instability by analysing the transverse motion. Its operational development is paramount as it is the only device that can observe the mode of instability (Fig. 4).

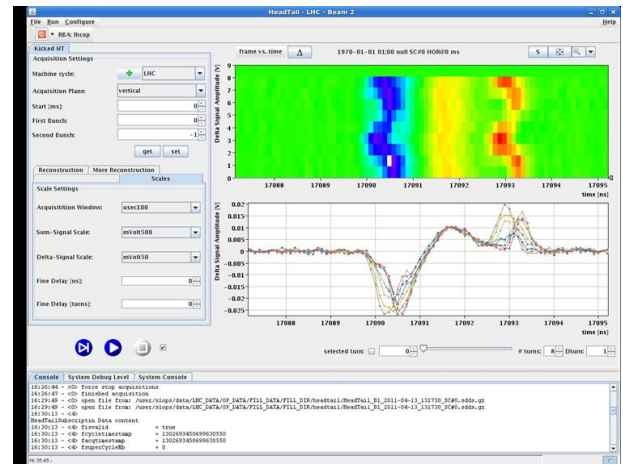

Figure 4: Head–tail monitor.

The tune measurement is vital for beam–beam studies. The BBQ system facilitated a reliable commissioning and operation of the LHC. Single bunch tunes could be measured by both BBQ and Schottky systems. During the 2012 proton run, the Schottky signal was reliable only on B1H for single and multi-bunch measurements at injection and stable beams. Unfortunately, large coherent signals saturated the pre-amplifiers in the other systems, rendering the signals unusable. The Schottky system is one of the systems capable of measuring bunch-by-bunch tune and chromaticity in a non-invasive, independent way. We strongly hope to have reliable tune and chromaticity measurements available after LS1. Schottky systems have been used successfully in other machines in the past (SPS, Tevatron, RHIC).

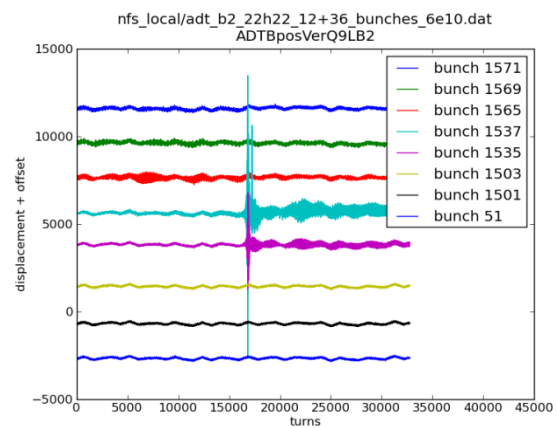

Figure 5: Instability at injection for different bunches during electron-cloud scrubbing studies, as measured by the transverse damper pickups. Only two bunches out of eight are affected by the instability.

The buffer on the transverse damper (ADT) beam position pickups can at present measure only up to 72 turns, but a higher number would be highly appreciated. If we could measure more than 72 turns, we could start to determine the frequency, mode pattern and rise time of the instability (see the example in Fig. 5, where the positions of only eight bunches were recorded to increase the available number of turns).

Chromaticity measurement and control is extremely important in a proton–proton collider. The LHC chromaticity should be nominally set to +2, but to guarantee this we need the chromaticity measurement to be precise within 1 unit to control this important machine parameter. The past LHC run with high-energy instabilities reinforced this need. The BBQ is the only instrument capable of measuring the chromaticity up to this moment, by means of an RF frequency modulation. The Schottky development should be encouraged to obtain reliable and continuous bunch-by-bunch tune and chromaticity measurements at any time to understand the instabilities issues and the differences between bunches (e.g. between Pacman bunches) [7] (Fig. 6).

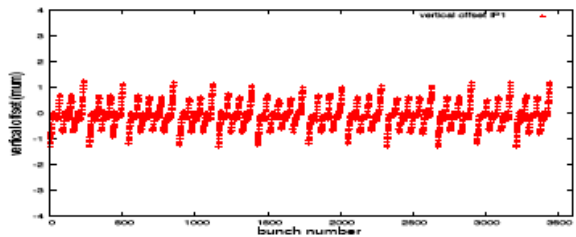

Figure 6: Vertical offset of the luminous region as measured by the ATLAS detector during a physics fill. The offset depends on the position of the bunch in the train [7].

The beam transfer function (BTF) could be useful at LHC to understand the tune spread. For example, at RHIC a BTF measurement is performed in stable beams every 15 minutes (Fig. 7). With this tool, we could study beam–beam coherent modes, normally Landau damped and not visible on BBQ measurement. We have enough simulation knowledge to start exploiting this technique. The LHC PLL could provide similar data after dedicated commissioning.

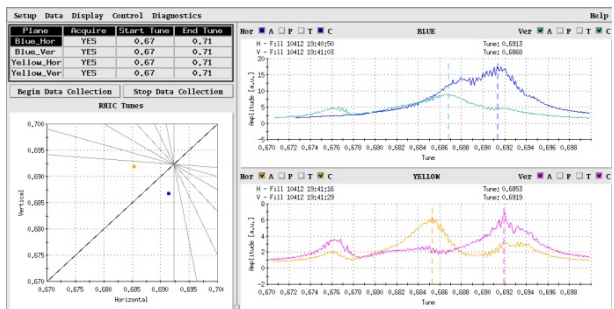

Figure 7: RHIC BTF measurements during stable beams.

Reliable, continuous, bunch-by-bunch emittance measurements are paramount to be able to study the beam–beam phenomena in the LHC.

In 2012. the BSRT (synchrotron light telescopes) was the only instrument capable of performing continuous, non-invasive, bunch-by-bunch measurements. Calibration was done with wire scanners. In the future we should improve the knowledge of the machine optics to reduce the uncertainty of this measurement. Fast scans have been available since May 2012 (1380 bunches can be scanned in 7 minutes) and are very helpful for day-to-day operation and machine studies. A new server that can perform the scans automatically has been available since October 2012.

## CONCLUSION

The beam–beam team would like to thank excellent collaboration with the Operation and Beam instrumentation groups. It allowed achievement of important results in the LHC beam–beam studies. To further investigate the origin of instabilities limiting the LHC performance, we would have to upgrade our instrumentation toward more bunch-by-bunch observation. These new tools will allow us to better control the LHC machine parameters, mitigate the instabilities, and optimize its operation and ultimately maximize the integrated luminosity.

LHC after LS1 will be a 'new' machine. After this important instrumentation upgrade we will be ready to face this new scenario.

## ACKNOWLEDGMENTS

The author would like to thank the following persons for their help in the preparation of this paper: OP group, BI group, ABP group, and in particular G. Arduini, X. Buffat, T. Baer, R. Calaga, D. Jacquet, R. Jacobsoon, R. Jones, W. Herr, W. Hofle, M. Lamont, G. Papotti, T. Pieloni, B. Salvant, R. Steinhagen, G. Trad, D. Valuch and J. Wenninger.